%% file: paper.tex
\documentclass[10pt,journal,compsoc]{IEEEtran}

\ifCLASSOPTIONcompsoc
\else
\fi
\ifCLASSINFOpdf
\else
\fi
\usepackage{tcolorbox}

\usepackage{booktabs} 
\usepackage[flushleft]{threeparttable}
\usepackage{xcolor}
\usepackage[utf8]{inputenc}
\usepackage{float}
\usepackage{amssymb}
\usepackage{ifthen}
\usepackage{multirow}
\usepackage{caption}
\usepackage{listings}
\usepackage{hyperref}
\hypersetup{colorlinks=true}
\usepackage{url,moreverb,xspace}
\usepackage{tcolorbox}
\usepackage{enumitem}
\usepackage{array,graphicx}
\usepackage{soul}
\usepackage{balance}
\usepackage{pifont}
\usepackage{etoolbox,siunitx}
\usepackage{nicefrac}
\usepackage{mathtools}
\usepackage{amsmath}
\usepackage{mathrsfs}

\newcommand{\dk}{Donkey Car\xspace} 

\newcommand{\davetwo}{\mbox{DAVE-2}\xspace} %

\newcommand{\changed}[1]{\textcolor{black}{#1}}

\newboolean{showcomments}
\setboolean{showcomments}{true}
\ifthenelse{\boolean{showcomments}}
{\newcommand{\nb}[2] {
  \fcolorbox{black}{gray!20}{\bfseries\sffamily\scriptsize#1:}
  {\sf\small$\blacktriangleright$\textit{#2}$\blacktriangleleft$}
}
}
{\newcommand{\nb}[2]{}
}

\newcommand{\head}[1]{\noindent\textbf{#1.}}

\begin{document}

%
\title{Mind the Gap! A Study on the Transferability of Virtual vs Physical-world Testing of Autonomous Driving Systems}
%
%
%
%

\author{Andrea~Stocco,~\IEEEmembership{Member,~IEEE Computer Society}, 
        Brian~Pulfer
        and~Paolo~Tonella,~\IEEEmembership{Member,~IEEE Computer Society}
\IEEEcompsocitemizethanks{%
\IEEEcompsocthanksitem Andrea Stocco and Paolo Tonella are with the Universit\`{a} della Svizzera italiana, Switzerland. \protect\\
\IEEEcompsocthanksitem Brian Pulfer is with the University of Geneva, Switzerland. The work has been carried out while at the Universit\`{a} della Svizzera italiana, Switzerland.}

}

%
%

\markboth{IEEE TRANSACTIONS ON SOFTWARE ENGINEERING,~Vol.~XY, No.~X, XYZ~2022}%
{Stocco \MakeLowercase{\textit{et al.}}: Mind the Gap! A Study on the Transferability of Virtual vs Physical-world Testing of Autonomous
Driving Systems}
%



\input{0-abstract}

\maketitle

\IEEEdisplaynontitleabstractindextext

%
\IEEEpeerreviewmaketitle

\input{1-introduction}

\input{2-background}
\input{3-testbed}

\input{4-approach}

\input{5-evaluation}

\input{6-discussion}
\input{7-related.tex}
\input{8-conclusion}
\input{acks}

\ifCLASSOPTIONcaptionsoff
  \newpage
\fi



\balance
\bibliographystyle{ieeetr}
\bibliography{paper}
\end{document}

%% file: 0-abstract.tex
\IEEEtitleabstractindextext{%
\begin{abstract}

Safe deployment of self-driving cars (SDC) necessitates thorough simulated and in-field testing. Most testing techniques consider virtualized SDCs within a simulation environment, whereas less effort has been directed towards assessing whether such techniques transfer to and are effective with a physical real-world vehicle. 

In this paper, we shed light on the problem of generalizing testing results obtained in a driving simulator to a physical platform and provide a characterization and quantification of the sim2real gap affecting SDC testing. 
In our empirical study, we compare SDC testing when deployed on a physical small-scale vehicle vs its digital twin. Due to the unavailability of  driving quality indicators from the physical platform, we use neural rendering to estimate them through visual odometry, hence allowing full comparability with the digital twin. 
Then, we investigate the transferability of behavior and failure exposure between virtual and real-world environments, targeting both unintended abnormal test data and intended adversarial examples. 
Our study shows that, despite the usage of a faithful digital twin, there are still critical shortcomings that contribute to the reality gap between the virtual and physical world, threatening existing testing solutions that only consider virtual SDCs. On the positive side, our results present the test configurations for which physical testing can be avoided, either because their outcome does transfer between virtual and physical environments, or because the uncertainty profiles in the simulator can help predict their outcome in the real world. 

\end{abstract}

\begin{IEEEkeywords}
AI Testing, Self-Driving Cars, Simulated Testing, Real-World Testing, Deep Neural Networks, Autonomous Vehicles.
\end{IEEEkeywords}}

%% file: 1-introduction.tex

\section{Introduction}\label{sec:introduction}

Self-driving cars (SDCs, hereafter) are automotive cyber-physical systems capable of sensing the environment and moving safely with little to no human driver input. Modern SDCs are realized with electric vehicles (EV) equipped with sensors such as cameras, LIDAR and GPS. Sensory data are processed by deep neural networks (DNNs) to generate predictions used to make driving decisions.
The consolidated industrial practice is to first collect real-world driving data that are recreated on high-fidelity and physically accurate simulators for large-scale testing. Then, the DNNs performance is assessed in real-world environments and vehicles, typically on private test tracks~\cite{Cerf:2018:CSC:3181977.3177753,10-million-miles,waymos-secret-testing,Thorn2018AFF}.

Existing DNN testing techniques from the software engineering literature, e.g., DeepXplore~\cite{deepxplore}, DeepTest~\cite{deeptest}, and DeepRoad~\cite{deeproad}, test the driving decision-making DNNs through \textit{model-level testing} (sometimes referred to as \textit{offline testing}~\cite{2020-Haq-ICST,Codevilla}), which means that the DNN (the model) is tested in isolation and on individual inputs, rather than considering the flow of inputs experienced in a simulation or in the real world. This does not take into account that each prediction (and subsequent driving decision) influences future events, which in turn influence future predictions and driving decisions. Thus, failures found by model-level testing (i.e., simple individual prediction errors) are hardly representative of real failures occurring when the DNN is embedded onto a vehicle~\cite{2020-Haq-ICST}. 

Specific, system-level, testing solutions for SDCs have been proposed~\cite{deeptest,deeproad,2021-Jahangirova-ICST,2020-Stocco-ICSE,10.1145/3449639.3459332,9438591,9376191,Gambi:2019:ATS:3293882.3330566}, in which the dominant approach for the empirical evaluation relies on the use of simulators, which does not guarantee that the observed failures, or lack thereof, correlate with those observable during on-road testing. Recent works have confirmed the need for real-world testing for robotic systems, as simulation platforms are often decoupled from the real world complexities~\cite{10.1145/3368089.3409743,AfzalSimulation21}.
As such, it would be important to ground the progress done in autonomous vehicle testing on both real-world environments and physical self-driving platforms.
Unfortunately, full-scale SDC testing presents severe time, space and cost constraints~\cite{2021-01-0248,bulsara2020obstacle}. Testing the limits of safety and performance on full-scale vehicles is costly, hazardous and outside the reach of most academic departments and research groups.

To mitigate these limitations, small-scale vehicles have emerged as an interesting alternative. Small-scale vehicles can be seen as an intermediate level before in-field, full-scale vehicles testing. 
Frameworks such as \dk~\cite{donkeycar}, AWS DeepRacer~\cite{aws-deepracer} or AutoRally~\cite{autorally} are adopted at the early stages of testing autonomous driving algorithms as they retain relevant photorealistic conditions of the driving environment which are experienced also by full-scale cars~\cite{2021-01-0248}. Moreover, these platforms are increasingly used by researchers who want to experiment their machine learning algorithms on real vehicles~\cite{9412011,viitala,DBLP:journals/corr/abs-1909-03467,DBLP:journals/corr/abs-1911-01562}, for educational purposes~\cite{autorally,8302386}, and research competitions~\cite{tum}. 

In this paper, we seize the opportunity of using small-scale vehicles to perform a comparative study between \textit{simulated and real-world testing of SDCs} and quantify the reality gap between testing environments. In our study, we leverage the \dk platform~\cite{donkeycar}, which allows developers to deploy a DNN that performs autonomous driving both on a 1:16 scale EV in a real test track as well as its virtualized counterpart. 
\changed{Several standards by ISO~\cite{iso26262} and NHTSA~\cite{Thorn2018AFF} mandate in-field testing to be implemented along with simulated testing to assess the risks associated with the real-world.}
\changed{Testing in a closed-track setting is one way to achieve such lifelike testing enviroments~\cite{Thorn2018AFF} and emerged as a consolidated industrial practice both for autonomous vehicles~\cite{Cerf:2018:CSC:3181977.3177753} and autonomous trucks~\cite{10.1007/978-3-030-59155-7_39}.}

\changed{
Our testbed is a closed environment with no traffic, where the vehicle drives at limited speed and the road layout is known. 
We tested the SDCs in a closed-loop environment because it allows us to remove confounding factors due to interfering tasks such as obstacle/pedestrian avoidance. Thus, our setting makes sure that the observed failures pertain to deficiencies of the DNN model that implements the lane-keeping task. 
This testbed is also in line with the scenes generated by test generators for lane-keeping DNNs~\cite{Gambi:2019:ATS:3293882.3330566,2020-Riccio-FSE}, which focus on the road shape of relatively short tracks to expose system's failures.}

We evaluate and test three state-of-the-art lane-keeping DNN models, i.e., NVIDIA's \davetwo~\cite{nvidia-dave2}, Chauffeur~\cite{chauffeur}, and Epoch~\cite{epoch}, through \textit{in-vitro} digital-controlled simulations and \textit{in-vivo} observations of deployed models in a real-world physical environment. We compared the testing results obtained in the virtual vs the real world, in terms of quality of driving and exposed failures. 
First, we assessed the transferability of driving behavior, using common driving quality metrics~\cite{2021-Jahangirova-ICST}. We found that steering angle predictions are statistically indistinguishable, whereas the SDCs' trajectories and the DNN models' uncertainties have non-negligible differences across the virtual and real-world environment.
Second, we studied failure exposure on 642 simulations under image corruptions and adversarial examples, finding that failures due to corruptions are more frequent in the physical environment and have higher severity, whereas SDCs in the real world are more robust to adversarial attacks.
Our analysis also revealed that models' uncertainty data from the virtual simulations allow halving the number of time- and cost-intensive physical tests.



The main contribution of this paper is the first empirical work that compares simulated vs real-world testing of SDCs at the system level, using a real-world EV. Our empirical study evaluates the transferability of driving behavior and failure exposure across environments. 
Practical take-away messages of our study are as follows:

\begin{itemize}

\item \changed{Concerning transferability of driving behaviours, despite individual DNN's predictions (i.e., steering angles) do transfer across virtual and real environments, offline testing alone is not a viable option. Indeed, SDC driving behaviour (as measured by lateral position and predictive uncertainty) is not always preserved across environments. Therefore, online testing of SDCs, possibly validated in-field, should be the de facto option.}

\item Concerning transferability of failures, our findings can help developers focus on the test conditions that are more likely to expose real-world failures and discard the conditions that are less promising. For example, virtual images are much easier to corrupt by adversarial attacks, but the associated failures do not correlate with the ones occurring in the real world. Thus, in-simulation adversarial attacks of virtual SDCs tend to overestimate the number of failures.

\item Our results suggest the usage of predictive uncertainty to develop test prioritization techniques for SDC that reduce physical testing, an important industrial need.

\item Neural image translation was effectively used for visual odometry to mitigate the sim2real gap. It could be adopted for the estimation of further sensory data, especially in domains in which in-field training is extremely costly (e.g., reinforcement learning~\cite{biagiola-plasticity}).

\end{itemize}


%% file: 2-background.tex
\section{Background}\label{sec:background}




%

\subsection{Model-level vs System-level Testing}
DNN models are typically tested by measuring either accuracy, mean squared error (MSE), mean absolute error (MAE), or other prediction metrics, on an unseen \textit{test set} of observations that were not used to train the model. 
We refer to this modality of testing as \textit{model-level testing}~\cite{2020-Riccio-EMSE}, because the model is tested as a standalone component, evaluating the predictions it makes on stationary datasets. This level of testing is comparable to unit testing for traditional software and is used by test engineers to reveal inadequately trained models~\cite{Humbatova:2020}. For DNN-based autonomous vehicles, model-level testing is ineffective in production. Studies~\cite{2020-Haq-ICST,Codevilla} have shown that  MSE is not a good estimator of the DNN's performance in the field, because it is not possible to identify and anticipate the causal chain from the individual, possibly small, prediction errors to failures. 

For this reason, in our work, we focus on \textit{system-level testing}, in which the DNN is embedded within a SDC to test the decision-making process, i.e., the effects that the predictions made by the DNN have on the behavior of the whole system. Thus, \textit{system failures} manifest as deviations from the expected system's behavior. For instance, safety requirements are violated when the vehicle drives off-road or causes harm to other vehicles, to the environment, or to people~\cite{2020-Stocco-ICSE}.

\subsection{Virtual vs Physical Testing Scenario}
System-level testing is often performed extensively within a simulation platform (e.g., CARLA~\cite{carla} or DeepDrive~\cite{dqm-deepdrive}), referred to as \textit{software-in-the-loop} (SIL), in which it is possible to measure, analyze, characterize, and reproduce driving failures safely. However, most simulators lack photo-realism and cannot reproduce real-world driving scenes with high fidelity and the question of whether the testing results would generalize to the real world remains open~\cite{AfzalSimulation21}. 

Thus, on-road system-level testing with a real physical vehicle---referred to as \textit{hardware-in-the-loop} (HIL)---remains quite important, despite being very expensive in terms of resources and time. For safety and logistic reasons, after simulated testing, SDC manufacturers perform on-road system-level testing on \textit{closed-loop private tracks}~\cite{waymo-driver}.
Only after the SDC is adequately tested within simulated and closed-loop tracks, the SDC is tested on \textit{public roads}~\cite{Cerf:2018:CSC:3181977.3177753,waymo-driver}. 

In our work, we study system-level testing of SDC models that perform \textit{behavioral cloning}, i.e., the autopilot learns the \textit{lane keeping} functionality from a dataset of driving scenes collected from a human driver.
Our focus is on the comparison between virtual and physical closed-loop track testing. In the rest of the paper, we shall use the terms virtual/digital/simulated, as well as the terms physical/real-world, interchangeably). 

%% file: 3-testbed.tex
\section{Testbed}\label{sec:donkeycar}

 
\subsection{\dk}\label{sec:donkey}

We use the Donkey Car\textsuperscript{\texttrademark} open-source framework~\cite{donkeycar} as the main testbed for our study. 
The framework includes open hardware to build 1:16 scale radio-controlled (RC) cars with self-driving capabilities, a Python framework that supports training and testing of SDCs that perform lane-keeping using supervised learning, and a simulator in which the real-world DonkeyCar is modeled with high fidelity, which allows to train and test the SDC without hardware.  

Among the RC platforms available on the market, only \dk fulfills our requirements for transferability studies between simulated and real-world testing of lane-keeping SDCs. For instance,  DeepPiCar~\cite{bechtel2018deeppicar} and JetRacer~\cite{jetracer}  lack a simulator, while the AWS DeepRacer~\cite{aws-deepracer} would restrict us to the usage of the AWS technology and is designed mostly to study RL-based autonomous vehicles, which is not the focus of this paper. Moreover, the \dk framework has already been used to experiment with autonomous driving software deployed on a real vehicle~\cite{2021-01-0248,9412011,viitala,DBLP:journals/corr/abs-1909-03467} and has an affordable cost.


\subsection{Framework and Simulation Platform}\label{sec:simulator}

The \dk framework provides Python scripts to drive the car manually, train DNN models, and deploy them.
The typical usage of the framework is as follows. First, training data are collected by manually driving the \dk through a web interface using a joystick or a virtual trackpad (we used a PS4\textsuperscript{\textregistered} controller). Each image is associated with the throttle and steering angle values that were observed when the frame was recorded.
Second, the framework supports the training of DNN models written in Keras/Tensorflow~\cite{chollet2015keras}. 
Finally, the newly trained model is evaluated on its capability of controlling the vehicle. 
Note that these steps can be carried out both in the real world and in the DonkeyCar simulator, developed on top of Unity~\cite{unity}. 

%% file: 4-approach.tex
\section{Metrics}\label{sec:metrics}

Our study requires collecting comparable driving profiles both in virtual and physical environments. 
We assess and compare the quality of driving by analyzing the distributions of three metrics, namely \textit{steering angle}, \textit{lateral deviation}, and \textit{predictive uncertainty}. The first two metrics are proposed by Jahangirova et al.~\cite{2021-Jahangirova-ICST} as effective metrics to evaluate the lane-keeping capability of SDC models. The last metric is used in the self-driving car domain to account for the DNN model's confidence~\cite{MichelmoreWLCGK20}. 

\subsection{Steering Angle} 

The main prediction computed by the DNN of the SDC consists of real numbers between $[-1,..., +1]$, where positive values indicate turning right, negative values indicate turning left, near-zero values indicate going straight. 

In our \textit{virtual} platform, a steering angle of +1 corresponds to 25$^{\circ}$, whereas a steering angle of -1 corresponds to -25$^{\circ}$. 
In the \textit{physical} Donkey Car, steering angles are converted into the vehicle's pulse lengths by an actuator called pulse-width modulation (PWM) servo motor. Prior to running experiments, we calibrated the PWM servo motor of the Donkey Car to match such angles. 
For our physical car, we found the values \texttt{STEERING\_LEFT\_PWM} $=$ 480 and \texttt{STEERING\_RIGHT\_PWM} $=$ 280 to correspond to $\pm$25$^{\circ}$.

\subsection{Lateral Position} 

In our setup, the car follows the middle line on a two-lane road (as if it were a single-lane, one-way road) and moves only forward.Thus, the lateral position of the vehicle with respect to the center of the road is the key telemetry value to assess the lane-keeping capability of our SDC models, as shown by a recent study~\cite{2021-Jahangirova-ICST}.

In our \textit{virtual} platform, we manually placed an ordered sequence of 22 waypoints onto the simulated track along the center of the road, to allow tracking the XTE of the vehicle during motion. Given the waypoints, the simulator measures the lateral position~\cite{2021-Jahangirova-ICST}, or cross-track error (XTE), as the distance between the car's cruising position and the segment between the two closest consecutive waypoints~\cite{2020-Stocco-GAUSS}.

In the \textit{physical} platform, XTE is not available from the physical vehicle as only camera images, steering angle, and throttle values are collected by the \dk. Given that manually labeling real-world data with dynamic telemetry data is an infeasible endeavour, we implemented a \textit{visual odometry} solution to estimate the lateral position by processing the onboard camera images~\cite{6096039} during the ego-motion of the physical SDC.


Our visual odometry solution uses neural style transfer and convolutional neural networks to build a telemetry data predictor that can operate on real-world driving images. We leverage the simulator to collect thousands of driving images paired with perfect telemetry (XTE) labels. However, an XTE predictor trained on such a dataset work on virtual, not real-world images. 
Thus, we train a generative adversarial network called CycleGAN~\cite{cyclegan} to create images that ``look like''  they were drawn from the real-world distribution. Then, we train a telemetry predictor that uses such generated data, labeled with the original labels from the simulator.
The use of CycleGAN~\cite{cyclegan} frees us from the constraint of collecting paired sets of images across environments, with the only requirement that they represent analogous driving images.

Our XTE predictor is a deep neural network that tackles a regression task, i.e., predicting a continuous value (e.g., the XTE) from a driving image. The DNN architecture is based on the pioneering work by Borjarski et al.~\cite{nvidia-dave2} on processing imagery data with convolutional neural networks. The network is composed of one normalization layer, five convolutional layers, and two fully connected layers. The first three convolutional layers use a 2$\times$2 stride and 5$\times$5 kernel; the last two convolutional layers are non-strided with a 3$\times$3 kernel.

We train our telemetry predictor on \textit{pseudo-real images} produced by CycleGAN, labeled with telemetry values obtained from the simulator for the simulated images (i.e., the CycleGAN's input). On the other hand, at prediction time the input to the predictor is a real-world image. Since the XTE predictor is quite important for our empirical study, we evaluate its accuracy with a dedicated RQ (\autoref{sec:evaluation}).

In both simulated and real-world environments, we use XTE to flag the car as off-road if the car deviates by more than half of the track's width (i.e., $\lvert XTE \rvert >$ 2.22, as $-2$ and $+2$ mark the lanes' borders; $0$ represents the middle of the lane, and $-1$, $+1$ are in-between).

\subsection{Predictive Uncertainty}

We measure predictive uncertainty with the variance of dropout-based DNNs' predictions, estimated using the Monte Carlo (MC) method, or MC-Dropout~\cite{10.5555/3045390.3045502}.  
MC-Dropout approximates epistemic uncertainty of DNNs that perform a regression, such as our SDC models~\cite{2020-Stocco-GAUSS,2021-Stocco-JSEP}. 

For the implementation of SDC models that support MC-Dropout, we followed the guidelines provided by Michelmore et al.~\cite{MichelmoreWLCGK20}. The dropout layers are enabled at both training and testing time (in Keras, this is realized by setting \texttt{training=True} in the Dropout layers). 
With dropout enabled at testing time, predictions are no longer deterministic: given the same driving image, the model could predict slightly different steering angle values every time the image is processed by the DNN. All such predictions are interpreted as a probability distribution and the value predicted by the DNN is the expected value (mean) of such probability distribution, whereas the variance of the observed probability distribution quantifies the uncertainty. A higher variance marks lower confidence, whereas a lower variance indicates higher confidence~\cite{10.5555/3045390.3045502}.

In our \textit{virtual} platform, we set the number of stochastic forward passes (i.e., the size of the sample used to estimate uncertainty) to match the batch size used during training (i.e., 64), whereas for the \textit{physical} Donkey Car, we used a batch size of 16 which allows processing at $\approx$20 Hz on a limited GPU memory (even smaller values, such as 10, are deemed reasonable~\cite{10.5555/3045390.3045502}).

%% file: 5-evaluation.tex
\section{Empirical Study}\label{sec:evaluation}

The \textit{goal} of the empirical evaluation is compare the driving behaviour and failure exposure of virtual vs physical SDCs across digital and real-world test environments.

\subsection{Research Questions}

We consider the following research questions:

\noindent
\textbf{RQ\textsubscript{0} (visual odometry accuracy):}
\textit{How accurate is visual odometry in estimating lateral position for autonomous vehicles driving in the real world?} 

In RQ\textsubscript{0} we compare the accuracy of lateral position estimations made by visual odometry with human-defined ground truth. The aim is to investigate whether this technique is accurate enough in the prediction of the lateral position metric on real-world data. This question serves as a pre-requisite for our empirical study and will assess whether the lateral position predictions by the visual odometry can be reliably used as visual surrogates of the true odometry in the real world. 

\noindent
\textbf{RQ\textsubscript{1} (transferability of behaviour):}
\textit{Do the behavioral and quality metric distributions observed in simulation transfer to the metric distributions estimated by visual odometry or measured in the real environment?}

In this RQ we compare the driving profiles observed in the virtual vs physical world. Ideally, good and faithful virtual environments should reproduce the same profiles that would be experienced in the real world.

\noindent
\textbf{RQ\textsubscript{2} (transferability of failures):}
\textit{What is the relation between system-level failures occurring in simulation and those observable in the real world? Can we expose the same real-world failures by running only the subset of tests that exhibit high telemetry (e.g., uncertainty) in the simulation?}

In this RQ we first compare the failures experienced during virtual vs physical system-level testing due to corrupted or adversarial settings. Ideally, good and faithful virtual environments should produce failures analogous to those experienced in the real world. 

\noindent
\textbf{RQ\textsubscript{3} (test selection):}
\textit{Can we minimize the cost of physical testing, in terms of reduction of real-world tests executed, thanks to model uncertainty profiles collected in the simulation platform?}

In this RQ we consider whether developers could avoid expensive real-world executions whose non-failing behavior can be predicted by looking at simulation telemetry data (i.e., steering angle, XTE, or uncertainty), focusing the testing budget on more promising test scenarios, i.e., those exhibiting high telemetry in simulation.

%



\subsection{Test Objects and Environment}\label{sec:alignment}

\subsubsection{Self Driving Car Models}\label{sec:models}
We test three existing DNN-based SDCs: NVIDIA's \davetwo~\cite{nvidia-dave2}, Chauffeur~\cite{chauffeur}, and Epoch~\cite{epoch}. 
We selected these models because they are robust, publicly available SDC models that have been objects of study in several testing works~\cite{deepxplore,deeptest,deeproad,2021-Jahangirova-ICST,2020-Stocco-ICSE}.

\davetwo consists of three convolutional layers, followed by five fully-connected layers.
Chauffeur uses a convolutional neural network to extract the features of input images, and a recurrent neural network to predict the steering angle from 100 previous consecutively extracted features.
Epoch is implemented as a simple convolutional neural network with three convolutional layers.

\subsubsection{Camera Calibration}
We matched all the parameters of the virtual camera in the simulator to those of the camera on the \dk (a Sony 8MP IMX219). Specifically, we set 160 degrees of field-of-view (FOV), an aperture of 2.35 (F), a focal length of 3.15mm, and a sensor size of 16mm. Finally, we set the frame rate in the simulator the same as in the physical camera (i.e., 21 fps), and we positioned the simulated camera to match the height of the real-world camera so that images are captured under the same perspective. 

For throttling, we set a constant throttle value of 0.2 in the simulator, corresponding to \texttt{THROTTLE\_FORWARD\_PWM} $= 500$ in the physical car, resulting in a maximum driving speed of $3.1$ mph ($5$ km/h, or $1.40$ m/s). 

\subsubsection{Testing Tracks}\label{sec:tracks}
We execute our experiments on the ``DIYRobocars Standard Track''~\cite{tracks}, 11 m long track which is 52 cm wide. Clockwise, the track features three curves on the right and one on the left. 
For the simulator, we developed a new Unity scene that resembles the real-world room and track. 
To perform this task, we imported a high-resolution picture of the real track into Unity. This ensures that the track used in the simulator is identical in road shape and colors to the one in the real world and that the proportions of the car with respect to the track are maintained.

\subsection{RQ\textsubscript{0}: Procedure and Metrics}

\subsubsection{CycleGAN Data Collection, Setup, and Training}

We trained a CycleGAN model on a custom dataset of 12,000 unpaired images, 5,361 virtual images and 6,639 real-world images, using the default hyper-parameters of the original paper~\cite{cyclegan}. \changed{Dataset's size was found adequate as it retains sufficient samples for representing our road track's shape.} 
Training images were collected with an original size of $320\!\times\!240$ pixels, cropped to size $320\!\times\!120$ to remove the background, and reshaped to $256\!\times\!256$ pixels, which is the default for CycleGAN. The output of the CycleGAN model is a $256\!\times\!256$ pixels image. 

The network was trained for 75 epochs with the default learning rate of $0.0002$, and for further 20 epochs featuring a linear decay of the learning rate towards zero. 
We visually assessed that the generated pseudo-real images highly resemble the real-world images (see \autoref{fig:example-cyclegan}). A detailed qualitative and quantitative evaluation of the outputs of generative networks is a challenging task with no consolidated solution~\cite{DBLP:journals/corr/abs-1802-03446}, and out of the scope of this paper. We rather evaluate the output of CycleGAN in an operative way, i.e., by using its outputs to train an XTE predictor and testing its accuracy on real-world data. The effectiveness of the XTE predictor would also implicitly validate the images translated by CycleGAN in their ability to retain all essential features needed for an accurate prediction.

\subsubsection{XTE Predictor Data Collection, Setup, and Training}

Training data were collected by instrumenting the \textit{virtual} DonkeyCar to drive for five laps in the simulator, following five different track trajectories along the track that would correspond to $XTE=[-2,-1,0,+1,+2]$. 

Overall, we obtained a balanced dataset of 5,362 training \textit{virtual} images labeled with accurate XTE values that we translated to the corresponding \textit{pseudo-real images} using CycleGAN. Such pseudo-real images produced by CycleGAN are labeled with XTE values obtained from the simulator for the corresponding simulated images (i.e., the CycleGAN's input). This means that labeled training data can be produced automatically, with no human input. Then, we trained the XTE predictor setting the maximum number of epochs to 100, with a learning rate of 0.0001. The network uses the Adam optimizer to minimize the mean squared error (MSE) between the predicted XTE value and the ground truth value. 

\subsubsection{Metrics for visual odometry evaluation}

For the purpose of answering RQ$_0$, 
we collected additional test data by manually driving the \textit{physical} DonkeyCar on the real-world track and by labeling manually the collected frames with ground truth XTE values. In fact, with RQ$_0$ we want to evaluate the accuracy of the XTE predictor when deployed on the real SDC, where the input is a real-world image.
To simplify frame capture, the speed was set to 0 mph and the car was moved from one position to the next manually. This was achieved by assigning the throttle's PWM a value below a certain configurable value (\texttt{THROTTLE\_FORWARD\_PWM} $<$ 480 in our setting). 
With this setting, the vehicle is regarded as ``active'', hence camera frame recording is enabled, even though the vehicle is not actually moving because the pulse width signal is too low to allow motion. This way we could capture frames of interest at different trajectories in a controllable way, by manually placing the vehicle at different locations on the track corresponding to the same five different trajectories of interest ($XTE=[-2,-1,0,+1,+2]$, the same values considered during the training set data collection). 

Overall, we obtained a dataset of 6,638 testing images, uniformly distributed over the five XTE classes. 
During testing, 
we evaluate the \textit{accuracy} of the XTE predictor by measuring the mean absolute error (MAE) both numerically and in centimeters on the manually-labeled test set of real-world images. 

\subsection{RQ\textsubscript{1}: Procedure and Metrics}


\subsubsection{SDC Data Collection}\label{sec:data-collection}

To obtain two comparable datasets of driving data, for each testing environment (virtual and physical world), we collected a training set of 30 laps by manually driving on both the virtual and physical tracks with a joystick. We incentivize the car to stay close to the centerline of the track. 

\subsubsection{SDC Model Setup \& Training}\label{sec:model-training}

For each DNN (\davetwo, Chauffeur, Epoch), we followed the guidelines by Bojarski et al.~\cite{nvidia-dave2} to train the SDC autopilots. We trained an individual SDC model on each training set (virtual and real-world), for a total of six models.
The upper bound number of epochs was set to 500, with a batch size of 64 and a learning rate of 0.0001. We used early stopping with a patience of 30 and a minimum loss change of 0.0005 on the validation set. The DNNs use the Adam optimizer~\cite{kingma2014adam} to minimize the MSE between the predicted steering angles and the ground truth value. We cropped the images to $140\times320$ by removing 100 pixels from the top to eliminate the above-horizon portion of the image, unnecessary for the lane-keeping task, and used data augmentation to mitigate the lack of image diversity in the training data using different image transformation techniques (e.g., translation). 
%



After training, we assessed that the six trained DNNs were able to drive in their corresponding testing tracks. We let each model drive for 50 consecutive laps and observed that they are able to drive without crashing or driving off-road ($\lvert XTE \rvert <$ 2.22 for all driving images). 
For the physical vehicle, we also carefully controlled the discharge of the \dk's battery and we recharged the battery if the voltage fell below 11.8V (12.7V being the maximum). Lower values were found to jeopardize the predictions and the quality of driving.

Prior to drawing conclusions across virtual and real-world models, we assessed whether the amount of available telemetry data was sufficient to characterize reliably each individual model. Specifically, we split the available telemetry data into two sets, one containing the driving data of the first 25 laps, and the second containing the driving data of the next 25 laps. Statistical power of the F-test for one factor ANOVA~\cite{cohen1988statistical} was always above the conventional threshold $\beta = 0.8$ 
and the $p$-value of the Mann-Whitney U test~\cite{Wilcoxon1945} was also above the conventional threshold $\alpha = 0.05$. This means that the telemetry data from the first 25 laps are statistically indistinguishable from those of the second 25 laps, both in the simulation and in the real environment. Thus, we can reliably use the metrics collected on such 50 laps for RQ\textsubscript{1}, i.e., transferability across environments. 

\subsubsection{Metrics for SDC Model Evaluation}\label{sec:model-testing}

We assess \textit{transferability} of the quality of driving by analyzing the distributions of three metrics, namely \textit{steering angle}, \textit{lateral deviation}, and \textit{predictive uncertainty} described in detail in \autoref{sec:metrics}. 
Both steering angle and XTE profiles characterize the behaviors of SDCs as erroneous steering angle predictions are likely to produce high positive or negative XTE values~\cite{2021-Jahangirova-ICST}. 
Uncertainty profiles do also characterize the driving behaviors from the point of view of the confidence (or lack thereof) of the DNN model while driving.
We compare the distributions of these three metrics in the virtual vs physical track, to investigate the transferability of the observed driving behaviors between the two environments. Ideally, all three metrics should have similar distributions in the simulation and in the real world.

\subsection{RQ\textsubscript{2}: Procedure and Metrics}\label{sec:rq3-proc}

\begin{figure}[t]
  \centering
  \includegraphics[trim=0cm 19cm 0cm 0cm, clip=true, width=\linewidth]{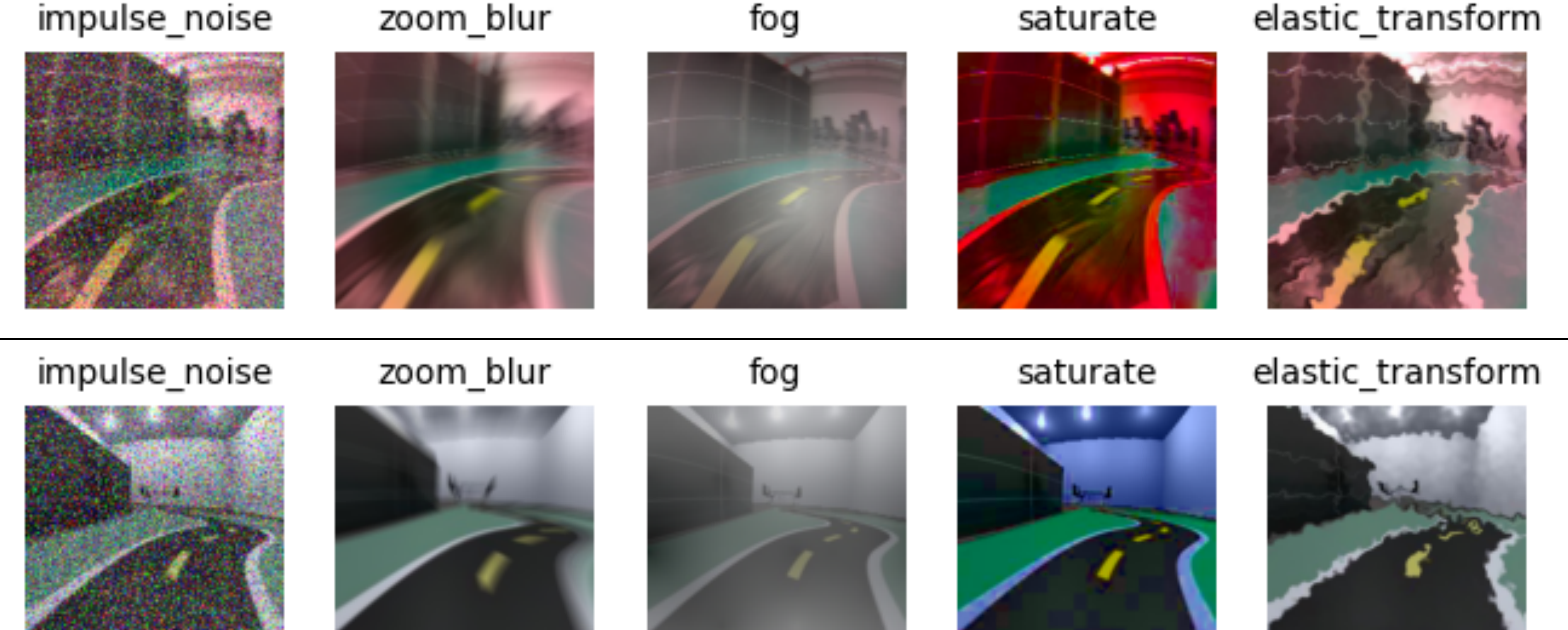}
  \caption{Corruptions on real-world/virtual driving images}
  \label{fig:corruptions}
\end{figure}

\subsubsection{Test Scenarios}
As our SDC models are constructed to be failure-free in nominal conditions (RQ\textsubscript{2}), with RQ\textsubscript{2} we test them by injecting unknown conditions (i.e., conditions different from those in the training set) onto the existing tracks \textit{in real-time during driving}. The main requirement is that such conditions should be applicable to virtual and physical settings alike.
We chose two different kinds of image perturbations, one black-box, and one white-box. 

Concerning the former, we test our SDCs under the image \textit{corruptions} by Hendrycks et al.~\cite{Hendrycks}, widely used to test DNNs that process imagery data. The paper proposes 19 corruptions belonging to five classes, namely \textit{noise} (four types: gaussian, shot, impulse, speckle), \textit{blur} (five types: gaussian, glass, defocus, motion, zoom), \textit{weather} (four types: fog, frost, snow, rain), \textit{luminance} (three types: contrast, brightness, saturate), and \textit{resolution reduction} (three types: JPEG compression, pixelate, elastic transform). Each corruption has a severity level from 1 to 5, indicating intensifying perturbations. 
\autoref{fig:corruptions} shows a few examples of simulated and real-world driving images.

\begin{figure}[t]
  \centering
  \includegraphics[trim=0cm 27cm 0cm 0cm, clip=true, width=\linewidth]{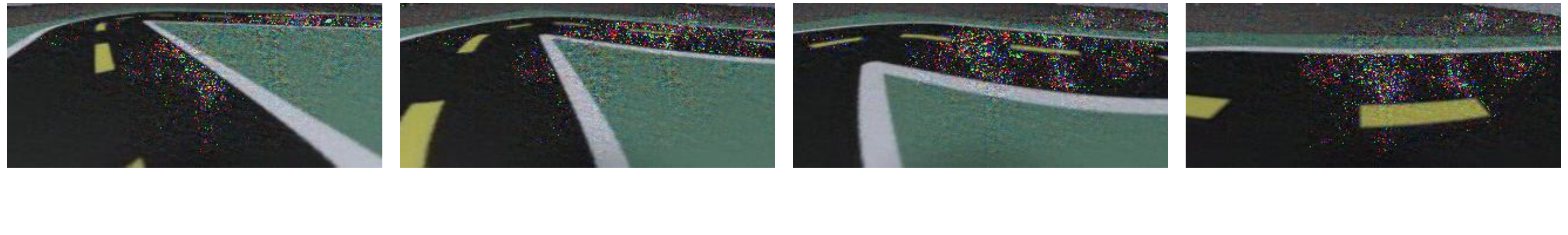}
  \caption{UAPr causing an OOT failure}
  \label{fig:adv-ex}
\end{figure}

For what concerns the white-box perturbations, we test our models against \textit{adversarial examples}. Unlike the corruptions by Hendrycks et al.~\cite{Hendrycks}, adversarial examples are minimal, synthetically crafted perturbations that cause a DNN to misbehave. In this paper we use the Universal Adversarial Perturbation on Regression (UAPr), 
a white-box targeted attack against end-to-end autonomous driving systems proposed by Wu and Ruan~\cite{adversarial-driving}. UAPr applies a universal perturbation to all driving frames of an autonomous vehicle, forcing it to drive in the desired direction (either left or right). This attack requires data-box access~\cite{2020-Riccio-EMSE} because the adversarial image is learned offline using the images of the training set of the DNN under test. The universal perturbation is generated through training by linearizing the output of the DNN until the minimum perturbation that changes the sign of the prediction to the desired direction (i.e., either steering left, or steering right) is found. During real-time execution, adversarial examples are generated by applying the generated universal perturbation to each input image by summation (see \autoref{fig:adv-ex}). 

We applied UAPr with two severity levels, either by adding the adversarial image as is or by doubling the magnitude of the perturbation it introduces. Moreover, we activated UAPr in the whole image stream starting from the beginning of the simulation (both for left and right), or only in the proximity of a curve (our testing track contains one left curve and three right curves), for a total of six configurations.  

Overall, we executed 642 one-lap simulations: 570 simulations enabling the corruptions (3 SDC models $\times$ 19 corruptions $\times$ 5 levels of severity $\times$ 2 testing environments) and 72 simulations enabling UAPr (3 SDC models $\times$ 6 configurations $\times$ 2 levels of severity $\times$ 2 testing environments).

\subsubsection{Metrics}
For each simulation, we classify its outcome using the following categorical scale: (1)~\textit{Successful (Succ.)} if the vehicle was able to travel the entirety of the track with no failures, (2)~\textit{OBE} (out-of-bound) failure, if the vehicle drove occasionally off-road while being able to recover back to the track and to complete the track, and (3)~\textit{OOT} (out-of-track) failure, if the vehicle drove off-road and was unable to recover back to the track. Clearly, OOT indicates a severe failure of the SDC, whereas OBE indicates a lower deviation from the main requirements. 

\begin{figure*}[t]
  \centering
\includegraphics[trim=0cm 26.5cm 0cm 0cm, clip=true, width=\linewidth]{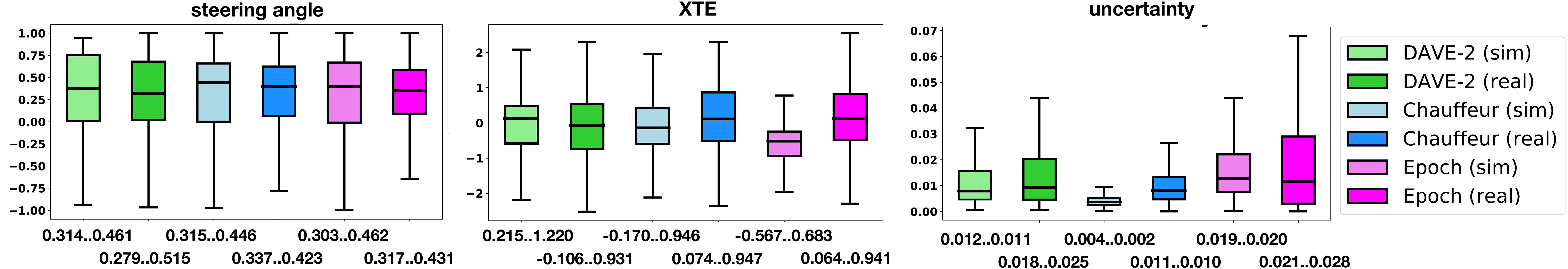}
  \caption{RQ\textsubscript{1}: distributions of telemetry and uncertainties in virtual vs physical test environments; x-axis: \texttt{mean..stddev}.}
  \label{fig:rq2}
\end{figure*}

Moreover, we divided our tracks into five distinct sectors and quantify the \textit{Severity} of a failure. Indeed, driving off-road at some initial part of the track indicates a worse driving capability than driving off-road near the track's end. 

\noindent
We calculate the severity of a failure as: 

\begin{equation}
	\textstyle{Severity} = \begin{cases}
			0, & \text{if Succ.} \\
			1 + (5 - \textstyle{TrackSector}), & \text{if OBE} \\ 
			10 + (5 - \textstyle{TrackSector}), & \text{if OOT}
	\end{cases} \nonumber
\end{equation}

\noindent
where \textit{TrackSector} is an integer in the range \numrange{0}{4}.
Thus, the \textit{Severity} metric ranges from 0 (i.e., no failure, the simulation is successful) to 15 (i.e., the worst case of failure, with an OOT in the first sector). 

\subsection{RQ\textsubscript{3}: Procedure and Metrics}\label{sec:rq4-proc}
\subsubsection{Procedure}
To analyze whether we can reduce the number of real-world tests to be executed, we collect driving telemetry data (steering angle, XTE, and uncertainty) under corrupted/adversarial settings (RQ\textsubscript{2}). Then, we consider the following cases: (1)~executions that are successful both in simulated and real environments (Succ. Sim $\rightsquigarrow$ Succ. Real), (2)~executions that are successful in simulation but expose OBE failures in the real environment (Succ. Sim $\rightsquigarrow$ OBE Real), and (3)~executions that are successful in simulation but expose OOT failures in the real environment (Succ. Sim $\rightsquigarrow$ OOT Real). 
We adopted a threshold corresponding to the median (50\textsuperscript{th} percentile) of the driving data distribution for all virtual driving executions. Only test scenarios having steering/XTE/uncertainty values above the threshold are selected for real-world execution.

To reduce the in-field testing cost,
among the test configurations that are successful in the simulation, we consider to execute only those that exhibit high telemetry (e.g., uncertainty above the 50\textsuperscript{th} percentile threshold) in simulation. The main hypothesis is that test selection can achieve a positive trade-off between the number of in-field tests that can be avoided (having low telemetry) vs the number of failures missed.

\subsubsection{Metrics}
For the selected test cases, we count the true positives (TP: selected tests that fail in the real environment with OBE or OOT), the false positives (FP: selected tests that do not fail in the real environment), the true negatives (TN: discarded that do not fail in the real environment), and the false negatives (FN: discarded that fail in the real environment with OBE or OOT). Ideally, good test selection should achieve high TP, TN, and low FP, FN values. We also measure the estimated saving as the fraction of non executed test cases: \textit{Saving} = (TN + FN) / (TP + TN + FP + FN).

\subsection{Results}

\subsubsection{Visual Odometry Accuracy (RQ\textsubscript{0})}
\begin{figure}[h!]
  \centering
  \includegraphics[trim=0cm 4cm 0cm 5.5cm, clip=true, width=0.6\linewidth]{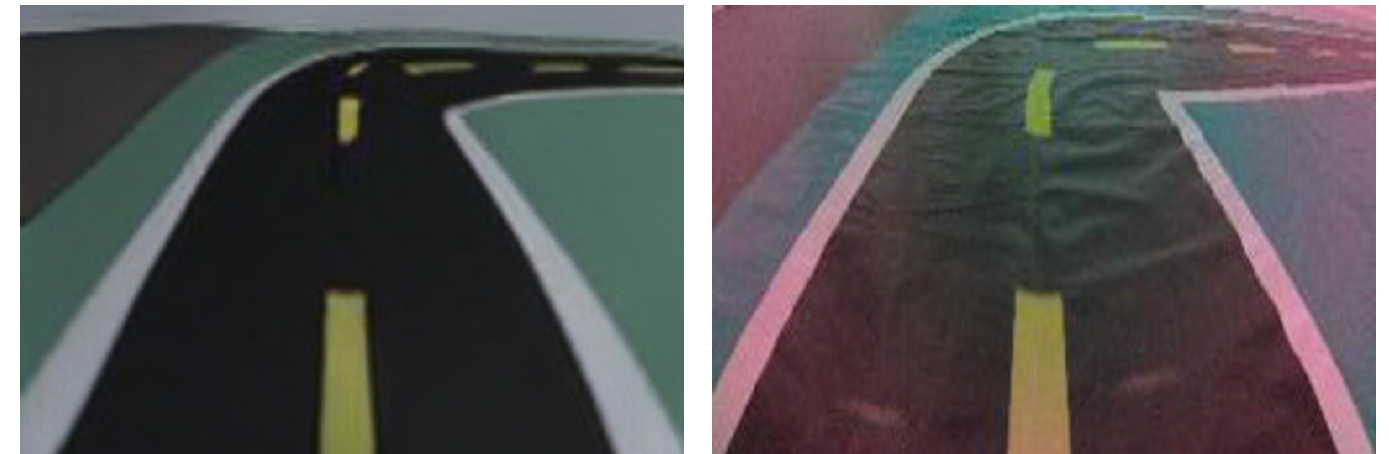}  
  \caption{Neural-generated real-world driving image (right), corresponding to a virtual, simulated image (left)}
  \label{fig:example-cyclegan}
\end{figure}

\autoref{tab:rq1} presents the accuracy results for visual odometry. Columns~2-3 report MAE values both numerically and in centimeters, whereas Columns~4 reports the MAE error as a percentage of the track width. The table reports the results on a per-class basis and the average across classes. Overall, visual odometry yields good results as a consequence of high quality translations (see \autoref{fig:example-cyclegan}).
The most interpretable results are those in centimeters, with an average error of 2.34 cm, corresponding to only 4\% of the road width.

\begin{center}
\begin{minipage}[t]{0.99\linewidth}
\textbf{$\Big \lceil$RQ\textsubscript{0}}: \textit{
The XTE estimation by visual odometry yields, on average, an error as small as 2.33 cm in the real-world, or 4\% of the road width. We conclude that we can use visual odometry as a reliable technique to estimate the lateral position of real-world images to study transferability of SDC behaviour.
\textbf{$\Big \rfloor$}}
\end{minipage}
\end{center}

\subsubsection{Transferability of Behaviours (RQ\textsubscript{1})}

\input{table-rq1}
\input{table-rq3}

\autoref{fig:rq2} compares the telemetry distributions for steering angle, XTE, and uncertainty for each SDC across environments. 
Statistical significance of the differences was assessed using the non-parametric Mann-Whitney U test~\cite{Wilcoxon1945} ($\alpha = 0.05$), and the Cohen's $d$ effect size~\cite{cohen1988statistical}.

Concerning steering angle distributions, no statistically significant differences were found during virtual and physical simulations across all models ($p$-value $> 0.05$; statistical power $> 0.8$). This is not entirely surprising, as the SDC models under test share the same architecture, the number of parameters, and training hyperparameters, whereas they mainly differ for the datasets they are trained on (yet representing the same track). Therefore, it is somehow expected that the set of predictions to navigate the tracks in nominal conditions is comparable across environments and corroborates our efforts in trying to align the two testing environments (see \autoref{sec:alignment}).

\begin{figure}[h!]
  \centering
  \includegraphics[trim=0cm 9cm 3cm 0cm, clip, width=\linewidth]{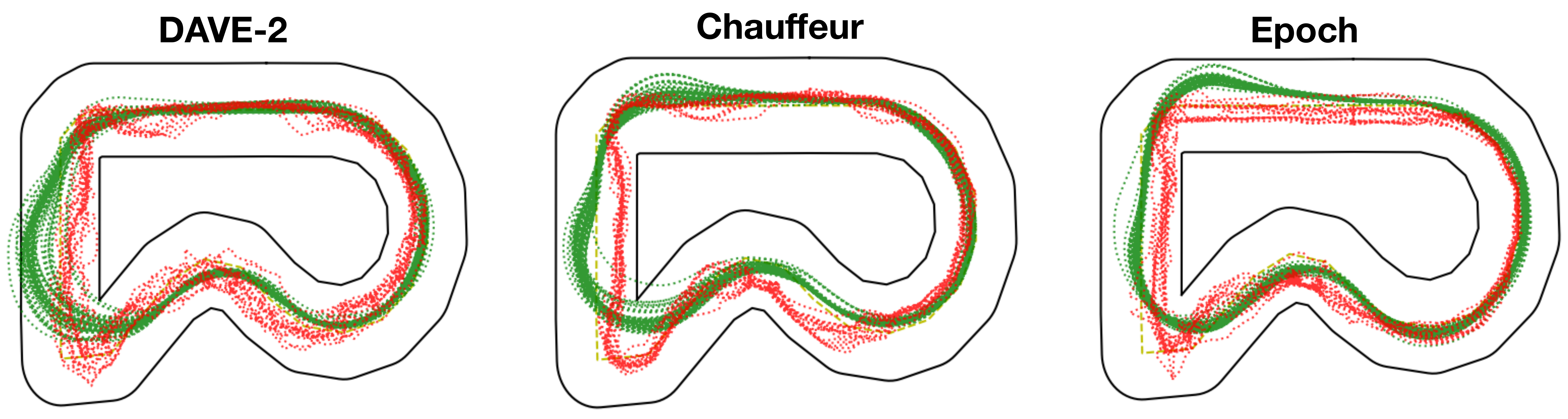}
  \caption{Virtual (green) and physical (red) trajectories}
  \label{fig:trajectories}
\end{figure}

Differently, XTE distributions are statistically different between virtual and physical simulations across all models ($p$-value $< 0.05$, with small/large/negligible effect sizes for \davetwo/Chauffeur/Epoch). 
\autoref{fig:trajectories} shows the trajectories of the SDCs. Trajectories by the physical \dk (red) are less smooth than those of the simulated virtual self-driving car (green) as they are characterized by sharper angles and more frequent steering adjustments. This may be due to the intrinsically higher difficulty of driving in the real world as compared to the simulator: real-world images are noisier and more difficult to interpret automatically for the DNN. Other explanations for the distinctness have to do with the physical actuators involved in real-world driving and the latency between the DNN predictions and the reactions of the car's mechanical components, which is absent in the simulator.

Our hypothesis is partially confirmed when looking at the model-level uncertainty distributions: we observed a statistically significant increase of the uncertainty values when predictions are made on the physical car ($p$-value $< 0.05$; negligible/small/large effect sizes for \davetwo/Chauffeur/Epoch). This means that the physical world introduces additional sources of uncertainty with respect to the simulated world due, for example, to ephemeral changes in illumination and noise, which may affect real images, while being absent in simulated images.

\begin{center}
\begin{minipage}[t]{0.99\linewidth}
\textbf{$\Big \lceil$RQ\textsubscript{1}}: \textit{The steering angle distributions observed in simulation do transfer to a physical SDC equipped with the same model. Differently, XTE values do not transfer when considering a physical vehicle. Likewise, uncertainty estimation does not transfer and uncertainty values of a physical SDC are generally higher than those exhibited by a simulated SDC, possibly because of sources of non-stationarities, uncertainty, noise and mechanical latencies affecting real world driving.$\Big \rfloor$}
\end{minipage}
\end{center}

\subsubsection{Transferability of Failures (RQ\textsubscript{2})}
\autoref{tab:rq3} presents the failures results separately for corruptions (aggregated by class) and adversarial examples. For each SDC model and for each testing environment, the table reports the number of configurations that are successful (Succ.), the number of those characterized by an OBE or by an OOT failure, and the severity score (Sever.). 

Concerning image corruptions, \davetwo is the best performing SDC model with the least number of failures, both in simulated and physical environments. 
The blur effect is the most severe corruption as far as OBEs are concerned for 4 out of 6 SDC models. 
Regarding OOTs, the weather effect is consistently responsible for OOTs in the virtual environment, whereas the blur effect is the main source for OOTs in the physical environment.
Overall, our results indicate an increasing number and severity of the failures due to corruptions in the real environment (for Severity: +37\% for \davetwo, +29\% for Chauffeur, and +2\% for Epoch).

Concerning adversarial attacks, \davetwo is the worst model, being unable to complete any virtual or physical simulation, while 
Chauffeur proved very robust to UAPr in the physical environment, with no failures exhibited by the \dk. 
The trend observed under adversarial attacks is opposite to the trend observed with corruptions: in the former case, the physical environment is more resilient to attacks than the simulator. The higher variability of real images, combined with the sources of noise in the real world, makes adversarial attacks less effective than on the clean, artificial images produced by the simulator, where pixel values are more predictable and hence more attackable.

\input{table-rq3-2}

In \autoref{tab:rq3-sev}, for each individual configuration (285 overall for corruptions and 36 for adversarial examples), we report the number of times in which the severity score was the same across environments, in which the severity score was worse for the virtual than for the physical environment, and vice-versa on a per-model basis and in total. 
Overall, we found that nearly 60\% of the configurations do generalize from simulated to real-world environments. For corruptions, the severity is usually amplified from virtual to physical environments, which means that we cannot completely avoid testing for corruptions in the physical environment.
Differently, no adversarial attacks on real-world images were found worse than the corresponding virtual ones. This essentially means that we can reliably test SDC under adversarial attacks in simulation platforms only, with a good likelihood of the results being transferable to a physical SDC with the same DNN model.

\begin{center}
\begin{minipage}[t]{0.99\linewidth}
\textbf{$\Big \lceil$RQ\textsubscript{2}}: \textit{System-level failures triggered by corruptions of the input images are more frequent in the real world than in simulation and tend to have higher severity. This may indicate that the simulator ignores real-world details that are important to expose failures. Interestingly, the opposite happens under adversarial attacks, that are less successful in the real world than in simulation. $\Big \rfloor$}
\end{minipage}
\end{center}


\input{table-rq4}

\subsubsection{Test Selection (RQ\textsubscript{3})}
Concerning real-world test reduction, we discarded the steering angle metric since values were found indistinguishable across environments (RQ\textsubscript{1}). For conciseness of presentation, we also do not show the XTE metric, as it was found ineffective for test selection. The only metric yielding promising results is predictive uncertainty. \autoref{tab:rq4} shows the results concerning the real-world test reduction achievable through the analysis of SDCs' predictive uncertainty during the virtual simulations. 
For each SDC model, \autoref{tab:rq4} reports the associated true/false positive/negative values, using the median as a threshold on the uncertainties of all virtual simulations. 
Results show a high reduction rate: for 40/61 configurations (66\%), physical testing on the \dk can be safely avoided (TN Succ. Real). At the same time, all 4 OBE failures (100\% of OBE Real) and a large fraction of OOTs (17/21, 81\% of OOT Real) are exposed in the real world, and only 4 OOT failures are missed. The expected time saving is quite high, as 44/86 (51\%) configurations do not need further real-world testing. In our setting, this saving amounts to $\approx$13 hours, considering an average of 5 min/test case execution in the physical environment.

\begin{figure}[t]
  \centering
  \includegraphics[trim=0cm 0.8cm 0cm 1.5cm, clip=true, width=0.8\linewidth]{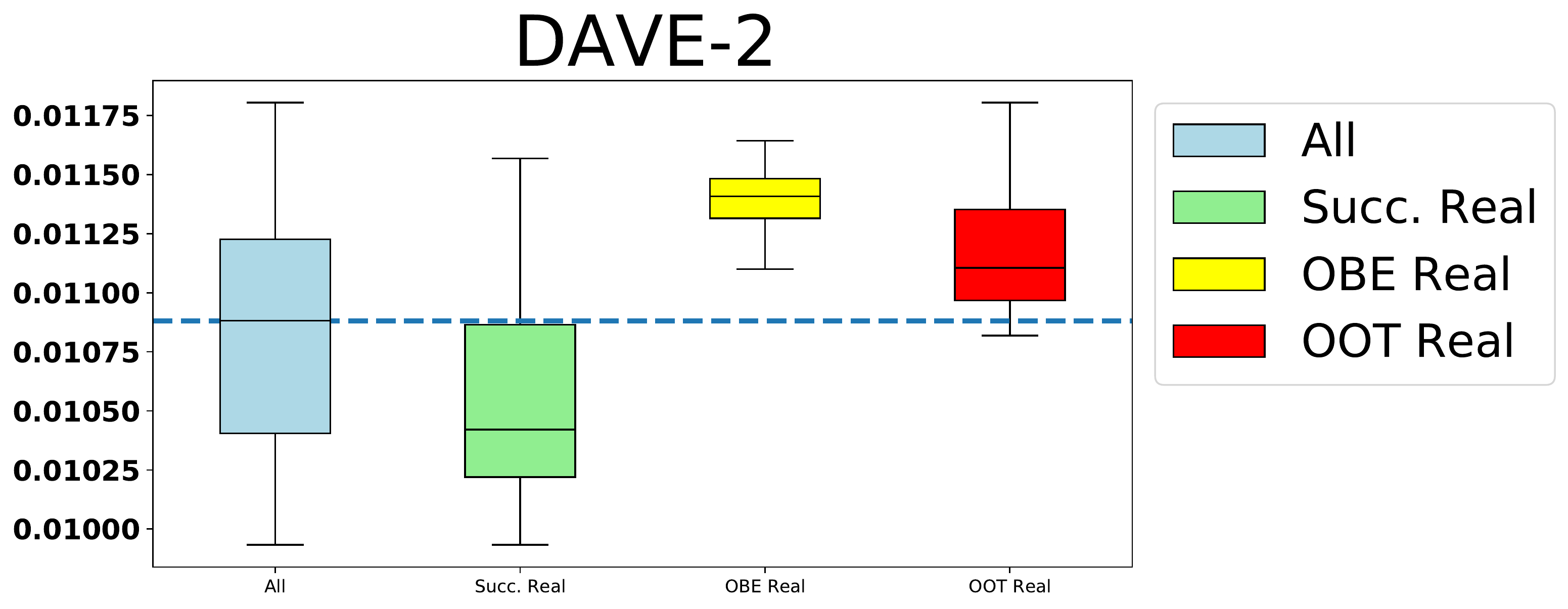}
  \caption{For \davetwo all test configuration with average uncertainty above the median are not executed in the real-world.}
  \label{fig:rq4-boxplot}
\end{figure}

\autoref{fig:rq4-boxplot} shows the uncertainty distributions for \davetwo, the only SDC model characterized by high failure variance across environments (see \autoref{tab:rq4}). For \davetwo, the median uncertainty threshold of 0.01088 provides a good cut-off value for the estimation of the real-world behavior.

\begin{center}
\begin{minipage}[t]{0.99\linewidth}
\textbf{$\Big \lceil$RQ\textsubscript{3}}: \textit{Model uncertainty allows to prioritize the successful simulations for real-world execution, providing an estimated time saving of around 51\%, countered by just 5\% missed OOT executions. $\Big \rfloor$}
\end{minipage}
\end{center}


\subsection{Threats to Validity}\label{sec:ttv}

\subsubsection{Internal validity}
We compared all SDCs under identical parameter settings.
One threat to internal validity concerns our custom implementation of the racing track within the simulator. However, this was unavoidable. We started from a picture of the real-world track to maintain its exact proportions and tried to reproduce the environment as faithfully as possible. 
Another threat may be due to our own data collection phase and training of SDCs, which may exhibit a large number of misbehaviors if trained inadequately or with poor quality data. We mitigated this threat by training and fine-tuning the best publicly available driving models, which were able to drive consistently in nominal conditions and were failing only in unseen conditions.

\subsubsection{External validity}
We used a limited number of SDC models in our evaluation, which poses a threat in terms of the generalizability of our results. We tried to mitigate this threat by choosing popular real-world SDC models which achieved competitive scores in the Udacity challenge. We considered only one physical track and only one 1/16 scale physical car. 
However, DonkeyCar was used as a proxy for full size SDCs also in previous studies~\cite{2021-01-0248,9412011,viitala,DBLP:journals/corr/abs-1909-03467}. 
Our results are obtained on a minimalistic scene, which was enough to investigate the lane keeping behaviour in isolation and to ensure statistical significance. We acknowledge that the availability of multiple and diverse physical tracks would be desirable. 
Other similar, small-scale platforms were found to retain key environmental characteristics experienced by full-scale SDCs~\cite{2021-01-0248}.
Hence, generalizability to other physical settings or RC platforms might not hold or may hold partially. Among the available RC platforms, only the \dk project fulfills all our requirements, namely a simulator, support from the community, and support for the supervised learning lane-keeping setting all at once. For instance, the DeepPiCar~\cite{bechtel2018deeppicar} and JetRacer~\cite{jetracer} projects lack a simulator, while the AWS DeepRacer~\cite{aws-deepracer} would restrict us to the usage of AWS technology. Moreover, AWS DeepRacer is designed mostly to study RL-based autonomous vehicles, which is not the focus of this paper.

\subsubsection{Reproducibility} 
The software artifacts, our results, demo videos, and the simulator are available~\cite{tool}. 
To \textit{replicate} our study, however, two open-source physical assets are needed, i.e., a DonkeyCar and the racing track ($\approx$500\$).

%% file: table-rq1.tex
\begin{table}[t]

\centering
\caption{RQ\textsubscript{0}: Errors in XTE prediction by visual odometry}
\label{tab:rq1}

\footnotesize

\setlength{\tabcolsep}{16pt}
\renewcommand{\arraystretch}{1}

\begin{tabular}{@{}lccc@{}}
\toprule
& \multicolumn{3}{c}{MAE}         \\ 

\cmidrule(l){2-4}

& \bf \#    & \bf cm    & \bf road width (\%) \\

\midrule

Outer lane (class 2)  & 0.26  & 2.98  & 5               \\
Left lane (class 1)  & 0.14  & 1.58  & 3               \\
Road center (class 0)  & 0.22  & 2.56  & 4               \\
Right lane (class -1) & 0.18  & 2.07  & 4               \\
Inner lane (class -2) & 0.21  & 2.49  & 4               \\

\midrule

Average      & 0.20 & 2.34 & 4               \\ 

\bottomrule

\end{tabular}
\end{table}

%% file: table-rq3.tex

\begin{table*}[t]
\scriptsize
\setlength{\tabcolsep}{1.1pt}
\renewcommand{\arraystretch}{1.2}

\caption{RQ\textsubscript{3}: Failures of SDC models for virtual and physical test environments}
\label{tab:rq3}

\begin{tabular}{@{}lllllllllllllllllllllllll@{}}

\toprule

& \multicolumn{8}{c}{\bf \davetwo} & \multicolumn{8}{c}{\bf Chauffeur} & \multicolumn{8}{c}{\bf Epoch} \\ 

\cmidrule(l){2-9}
\cmidrule(l){10-17}
\cmidrule(l){18-25} 

& \multicolumn{4}{c}{\textit{sim}} & \multicolumn{4}{c}{\textit{real}} & \multicolumn{4}{c}{\textit{sim}} & \multicolumn{4}{c}{\textit{real}} & \multicolumn{4}{c}{\textit{sim}} & \multicolumn{4}{c}{\textit{real}} \\ 

\cmidrule(l){2-5}
\cmidrule(l){6-9}
\cmidrule(l){10-13} 
\cmidrule(l){14-17} 
\cmidrule(l){18-21} 
\cmidrule(l){22-25} 

& Succ. & OBE & OOT & Sever. & Succ. & OBE & OOT & Sever. & Succ. & OBE & OOT & Sever. & Succ. & OBE & OOT & Sever. & Succ. & OBE & OOT & Sever. & Succ. & OBE & OOT & Sever. \\

\midrule

\textbf{\textit{Corruptions}} &  &  &  &  &  &  &  &  &  &  &  &  &  &  &  &  &  &  &  &  &  &  &  &  \\
\quad All & 38 & 16 & 41 & \bf 6.0 & 29 & 9 & 57 & \bf 8.2 & 32 & 17 & 46 & \bf 6.9 & 27 & 2 & 66 & \bf 8.9 & 18 & 12 & 65 & \bf 9.2 & 23 & 4 & 68 & \bf 9.4 \\
\qquad Noise & 5 & 4 & 11 & 7.5 & 7 & 1 & 12 & 8.1 & 8 & 2 & 10 & 6.6 & 8 & 1 & 11 & 7.3 & 2 & 0 & 18 & 12.5 & 2 & 0 & 18 & 12.5 \\
\qquad Blur & 15 & 3 & 7 & 4.2 & 7 & 3 & 15 & 8.5 & 6 & 8 & 11 & 7.4 & 4 & 0 & 21 & 10.9 & 0 & 7 & 18 & 10.2 & 2 & 2 & 21 & 11.8 \\
\qquad Weather & 1 & 4 & 15 & 10.7 & 4 & 2 & 14 & 9.9 & 2 & 2 & 16 & 11.2 & 2 & 1 & 17 & 11.1 & 1 & 0 & 19 & 13.1 & 1 & 1 & 18 & 12.4 \\
\qquad Luminance & 7 & 2 & 6 & 5.4 & 1 & 1 & 13 & 11.2 & 6 & 2 & 7 & 7.0 & 1 & 0 & 14 & 12.6 & 5 & 2 & 8 & 7.9 & 5 & 1 & 9 & 8.5 \\
\qquad Resolution Reduction & 10 & 3 & 2 & 2.3 & 10 & 2 & 3 & 3.3 & 10 & 3 & 2 & 2.3 & 12 & 0 & 3 & 2.6 & 10 & 3 & 2 & 2.3 & 13 & 0 & 2 & 1.7 \\ [0.5em]

\textbf{\textit{Adversarial Examples}} &  &  &  &  &  &  &  &  &  &  &  &  &  &  &  &  &  &  &  &  &  &  &  &  \\
\quad All & 0 & 0 & 12 & \bf15.0 & 0 & 0 & 12 & \bf15.0 & 4 & 4 & 4 & \bf5.1 & 12 & 0 & 0 & \bf0.0 & 0 & 0 & 12 & \bf13.8 & 2 & 0 & 10 & \bf8.1 \\
\qquad No Steer Left & 0 & 0 & 4 & 15.0 & 0 & 0 & 4 & 15.0 & 0 & 4 & 0 & 4.0 & 4 & 0 & 0 & 0.0 & 0 & 0 & 4 & 13.5 & 2 & 0 & 2 & 5.5 \\
\qquad No Steer Right & 0 & 0 & 8 & 15.0 & 0 & 0 & 8 & 15.0 & 4 & 0 & 4 & 6.3 & 8 & 0 & 0 & 0.0 & 0 & 0 & 8 & 14.0 & 0 & 0 & 8 & 10.8 \\

\bottomrule
\end{tabular}
\end{table*}

%% file: table-rq3-2.tex

\begin{table}[t]
\centering
\footnotesize
\setlength{\tabcolsep}{5.7pt}
\renewcommand{\arraystretch}{1}
\caption{Failure matches across virtual and physical tests.}
\label{tab:rq3-sev}
\begin{tabular}{@{}lcccccccc@{}}
\toprule

& \multicolumn{2}{c}{\bf \davetwo} & \multicolumn{2}{c}{\bf Chauffeur} & \multicolumn{2}{c}{\bf Epoch} & \multicolumn{2}{c}{\bf Total} \\ 

\cmidrule(l){2-3}
\cmidrule(l){4-5}
\cmidrule(l){6-7}
\cmidrule(l){8-9}

& \# & \% & \# & \% & \# & \% & \# & \% \\

\midrule

\textbf{\textit{Corruptions}}  & \multicolumn{1}{l}{} & \multicolumn{1}{l}{} & \multicolumn{1}{l}{} & \multicolumn{1}{l}{} & \multicolumn{1}{l}{} & \multicolumn{1}{l}{} & \multicolumn{1}{l}{} & \multicolumn{1}{l}{} \\

\quad sim \textcolor{green}{\ding{52}} real \textcolor{green}{\ding{52}} & 49 & 17 & 48 & 17 & 68 & 24 & 165 & \bf 58 \\
\quad sim \textcolor{red}{\ding{56}} real \textcolor{green}{\ding{52}} & 10 & 4 & 18 & 6 & 16 & 6 & 44 & \bf 15 \\
\quad sim \textcolor{green}{\ding{52}} real  \textcolor{red}{\ding{56}}  & 36 & 13 & 29 & 10 & 11 & 4 & 76 & \bf 27 \\ [0.5em]

\textbf{\textit{Adversarial Examples}} & \multicolumn{1}{l}{} & \multicolumn{1}{l}{} & \multicolumn{1}{l}{} & \multicolumn{1}{l}{} & \multicolumn{1}{l}{} & \multicolumn{1}{l}{} & \multicolumn{1}{l}{} & \multicolumn{1}{l}{} \\
\quad sim \textcolor{green}{\ding{52}} real \textcolor{green}{\ding{52}} & 12 & 100 & 4 & 33 & 4 & 8 & 20 & \bf 56 \\
\quad sim \textcolor{red}{\ding{56}} real \textcolor{green}{\ding{52}} & 0 & 0 & 8 & 67 & 8 & 4 & 16 & \bf 44 \\
\quad sim \textcolor{green}{\ding{52}} real  \textcolor{red}{\ding{56}} & 0 & 0 & 0 & 0 & 0 & 0 & 0 & \bf 0 \\ 

\bottomrule

\end{tabular}
\end{table}

%% file: table-rq4.tex

\begin{table}[b]
\centering
\footnotesize
\setlength{\tabcolsep}{2pt}
\renewcommand{\arraystretch}{1}
\caption{Test selection based on uncertainty for configurations above the median: true/false positives/negatives}
\label{tab:rq4}
\begin{tabular}{@{}lcccccccccc@{}}

\toprule

{\bf Succ. Sim $\rightsquigarrow$} & \multicolumn{2}{c}{\bf Succ. Real} & \multicolumn{2}{c}{\bf OBE Real} & \multicolumn{2}{c}{\bf OOT Real} & \multicolumn{1}{c}{\bf Tot. Sim.} & \multicolumn{3}{c}{\bf Saving} \\ 

\cmidrule(l){2-3}
\cmidrule(l){4-5}
\cmidrule(l){6-7}
\cmidrule(l){8-8}
\cmidrule(l){9-11}

& FP & TN & TP & FN & TP & FN & \# & \# & \% & hr  \\
\midrule
DAVE-2 & 4 & 17 & 4 & 0 & 10 & 2 & 37 & 19 & 51 & 4.28 \\
Chauffeur & 9 & 13 & - & - & 6 & 2 & 30 & 15 & 50 & 4.17 \\
Epoch & 8 & 10 & - & - & 1 & 0 & 19 & 10 & 53 & 4.39 \\
\midrule
Total & 21 & 40 & 4 & 0 & 17 & 4 & 86 & \bf 44 & \bf 51 & \bf 12.83 \\
\bottomrule
\end{tabular}
\end{table}

%% file: 6-discussion.tex
\section{Discussion}\label{sec:discussion}

\head{Neural translation is effective to mitigate the sim2real gap} 
The CycleGAN used for visual odometry (RQ\textsubscript{0}) could find applications that go beyond real-world metric estimation.
Previous work~\cite{2020-Haq-ICST} comparing offline vs online testing suffers from the problem of aligning labeled data from the two environments (real and simulated world), as the offline training datasets of real-world driving images with steering angle labels (e.g., by Udacity's~\cite{udacity-datasets} or Waymo's~\cite{waymo-datasets}) might be difficult to pair with labeled simulated images obtained from a simulator. Such alignment was achieved heuristically~\cite{2020-Haq-ICST}; with CycleGAN, developers and researchers could translate real images to pseudo-simulated ones, and then find the closest labeled image from the simulated dataset, according to some image-specific distance metric.
CycleGANs could be useful in the context of reinforcement learning-based autonomous driving~\cite{biagiola-plasticity,pmlr-v123-o-kelly20a}, as the estimated XTE could provide the necessary reward signal during training. Indeed, our visual odometry approach could be utilized to estimate XTE more accurately than existing computer vision heuristics that are hard to tune~\cite{kiran2021deep}.

Concerning the usage of real-world driving scenes to inform the driving in simulation, we envision two main applications. First, CycleGAN could be used for producing photo-realistic training data that could help the training of SDC in the simulator but with pseudo-real data gathered from on-road testing. 
Second, CycleGAN could be also useful to perform system-level virtual testing on pseudo-real testing environments that are dynamically generated and loaded at runtime as the simulated vehicle executes.

\head{Model-level metrics do transfer, system-level metrics do not}
The contribution of our telemetry predictor was instrumental to the correct assessment of the sim2real gap. 
Without the XTE estimations made by our technique, we could study the transferability only using the steering angle information, from which we would wrongly conclude that transferability is high (distributions of steering angles are not statistically different across environments, see our RQ\textsubscript{1} results). Differently, using the XTE estimations, we could observe that, despite the steering angles being distributed similarly, the trajectories taken by the same model in the two environments differ substantially. Analysis of the uncertainty profiles offers an explanation for such differences. Thus, SDC driving models may have different driving behaviors that are only exposed by system-level testing, and not at the model level. This is in line with the findings reported by Ul Haq et al.~\cite{2020-Haq-ICST} on the low level of agreement between model and system level testing of SDC. 

\head{Only specific types of virtual failures transfer to the real world}
For a good number of conditions ($\approx$60\%), testing results do generalize across environments (RQ\textsubscript{2}). This may be due to a faithful digital recreation of the real-world testing environment, necessary to fully exploit simulated testing, as suggested by industrial practices~\cite{10-million-miles,waymos-secret-testing,waymo-driver}. As a practical message, researchers can pre-select the testing conditions for which virtual and real-world failures correlate. For instance, our study revealed that adversarial attacks of virtual DNNs tend to overestimate the number of failures. 

\head{Predictive uncertainty can help prioritize test cases}
For the mismatching cases, we have shown that MC-Dropout uncertainty on the virtual platform allows the selection of the most promising test cases that need to be executed in the real world (RQ\textsubscript{3}). 
This is in line with the results of the study by Weiss and Tonella~\cite{Weiss2022SimpleTW} in which model uncertainty was shown to be a good metric for test input prioritization of DNN classifiers.
The potential saving for our rather short track (\mbox{-12.8 hours}) may be amplified substantially on longer tracks in which tests unlikely to expose failures in the real world represent a major cost.  

\head{Real-world physical testing is still needed}
Our results quantify the gap between the testing results from \textit{in-vitro} simulated experiments and the real-world \textit{in-vivo} observations of deployed models. We hypothesize the main root causes for such a gap to be the poor photo-realism and inadequate representation of the vehicle's sensory and environmental uncertainty sources.  
In the physical car, driving is affected by inevitable real-time stochasticities that are immaterial in the virtual world. 
For instance, the throttle can be adversely affected by the friction of the surface and battery voltage, whereas steering angle prediction can be affected by noise such as sudden spikes in luminosity, latency between prediction and actuation, and other forms of disturbances that are unveiled only in a physical setting. 
Given the current state of the practice, a proper amount of testing in physical conditions is still necessary to help detect and correctly handle discrepancies between the virtual and real environments. To reduce such disagreements, we suggest using an array of redundant digital twins that exhibit complementary performance and mitigate the weaknesses of the single-simulator approach, in line with the suggestions reported by Borg et al.~\cite{borg-digital-twin} and Arrieta~\cite{DBLP:journals/corr/abs-2101-05697}.  

%% file: 7-related.tex
\section{Related Work}\label{sec:relwork}

\subsection{Model- and System-level Testing Approaches}
Most approaches use \textit{model-level testing} to test DNN autopilots under corrupted images~\cite{deepxplore,deeptest} or GAN-generated driving scenarios~\cite{deeproad} without however testing the self-driving software on its target environment. Dang et al.~\cite{deng2020analysis} study the robustness of DNN driving models against different adversarial attacks. 
While we also use (universal) adversarial attacks, 
differently, in our work, we focus on system-level testing and on simulated vs real SDCs, finding simulated SDCs generally more susceptible to adversarial attacks than their real-world counterparts when tested at the system level. 
Concerning \textit{system-level testing} techniques for AVs, researchers proposed techniques to generate scenarios that cause AVs to misbehave~\cite{2020-Stocco-ICSE,Gambi:2019:ATS:3293882.3330566,2020-Riccio-FSE,2021-Stocco-JSEP,2022-Stocco-ASE}. 
These works only consider simulated testing, whereas we compared virtual vs physical environments. The applicability of test generators to real driving scenarios remains unexplored. 

\subsection{Empirical Studies}

Recent works have confirmed the need for real-world testing of robotic systems, as simulation platforms are often decoupled from the real-world complexities~\cite{10.1145/3368089.3409743,AfzalSimulation21}. 
Our work is the first one to quantify such a reality gap and confirms the existing anecdotal knowledge in the field of autonomous driving. 
Other works investigate the relation between model-level vs system-level testing metrics within a simulation environment. Codevilla et al.~\cite{Codevilla} found that offline prediction errors are not correlated with driving quality, and two DNN models with analogous error prediction rates may differ substantially in their driving quality. 
Haq et al.~\cite{2020-Haq-ICST} compare the distributions of offline vs online predictions, but they consider only simulated environments (i.e., virtual SDCs) for the online predictions. They also compare offline and online errors, finding offline testing more ``optimistic'' than online simulation, as the accumulation of small, acceptable offline errors might result in safety online violations, during the simulation.
In our empirical work, we focused on the difference between simulated and real-world environments, rather than offline vs online testing. We found a measurable gap between virtual and physical environments during system-level testing, which questions the transferability between simulation and real environment, a previously unexplored topic. 
 



\subsection{Virtual vs Physical Testing of Autonomous Vehicles}
Chen et al.~\cite{8723561} embed a real hardware control unit within a simulation platform to verify the validity of self-driving DNNs in virtual scenes, including perception, planning, decision-making, and control. 
Hildebrandt and Elbaum~\cite{9561240} address the reality gap by integrating sensory data from simulation and the real world to provide the autonomous system (a drone) with mixed reality. 
Differently, we do not hybridize simulation and real-world testing, as our goal is to understand the transferability between the two. 

Researchers have been using \dk~\cite{9412011,viitala,DBLP:journals/corr/abs-1909-03467} or other frameworks~\cite{DBLP:journals/corr/abs-1911-01562,pmlr-v123-o-kelly20a} to study reinforcement learning algorithms for autonomous driving. 
Verma et al.~\cite{2021-01-0248} compare different scaled vehicles concluding that such platforms allow the rapid exploration of many different test tracks while retaining realistic environmental conditions, which provides further justification for our choice to use \dk.
The usefulness of scaled SDC is even increased for fields in which fleets of vehicles are needed. Hyldmar et al.~\cite{Hyldmar} use scaled cars to study cooperative driving experiments and autonomous control strategies.
Scaled Jetson Nano-based vehicles have been used also to study multi-agent coordination planning for multi-goal tasks~\cite{Srinivasa}. 
Kannapiran and Berman~\cite{9341770} use scaled vehicles to study the interactions between human-driven and self-driving vehicles in a safe, controlled environment. 
Differently, in our paper, we use scaled vehicles to understand the transferability of testing results for SDCs. 

\subsection{Digital Twins}
Researchers aim to reproduce real-world conditions within a simulation environment using digital twins~\cite{10.1007/978-3-030-59155-7_39,9369807,san2021digital,9392784,9062813}. 
Yun et al.~\cite{9369807} use the GTA5 videogame to test an object recognition system. 
Barosan et al.~\cite{10.1007/978-3-030-59155-7_39} describe a digital twin for an autonomous truck operating in a distribution center. Similarly to our scenario, the scene is a closed environment with no traffic, where the vehicles drive at relatively low speeds and the road layout is known. Differently, no testing was performed using the digital twin to assess the faithfulness of the simulator at reproducing real-world failures. 
Ayerdi et al.~\cite{Ayerdi-TREL} use metamorphic relations to test an autonomous driving system in a simulator. 
Borg et al.~\cite{borg-digital-twin} report low agreement between testing results across (two) simulation platforms but without any assessment of the physical counterpart. 
Differently, in our paper, we investigate transferability and differences between a digital twin and a physical SDC. 

\subsection{GAN-based Testing of Autonomous Vehicles}

DeepRoad~\cite{deeproad} uses UNIT~\cite{unit} to generate accurate photo-realistic paired driving scenes for SDC testing, which were evaluated for their capability of exposing individual prediction errors. 
Kong et al.~\cite{physgan} generate realistic adversarial billboards within real-world images that can confound the vehicle. In our work, we also use universal adversarial perturbation at the system level, finding comparable results in terms of virtual/physical robustness.
DeepBillboard~\cite{deepbillboard} is among the first studies to generate physical-world tests for autonomous driving systems, using both digital and physical adversarial perturbation generation for impacting offline steering decisions. DeepBillboard generates perturbations of real-world images but uses offline testing (i.e., no actual driving, only individual DNN predictions). In our work, we apply perturbations of real-world images using online testing, i.e., at runtime, during the motion of the vehicle, which is the only way to evaluate system-level failures.
SilGAN~\cite{DBLP:journals/corr/abs-2107-07364} uses GANs to generate driving maneuvers for software-in-the-loop testing. SurfelGAN~\cite{DBLP:journals/corr/abs-2005-03844} is a technique developed at Waymo to generate realistic sensor data for autonomous driving simulation without requiring the manual creation of virtual environments and objects. 

Differently from existing works, we use CycleGAN, which requires no pairing, to generate pseudo-real driving scenes that can inform a telemetry predictor to support virtual vs physical testing. Our results show that this contribution was indispensable for the correct assessment of the sim2real gap.

%% file: 8-conclusion.tex
\section{Conclusions}\label{sec:conclusions}

Virtual testing of autonomous vehicles with simulators is widely adopted. However, the lack of adequate fidelity and environmental modeling threatens the generalizability of testing results to the real world.
While the development of scalable, physically accurate, and highly photo-realistic simulators is a cornerstone for autonomous vehicle development, real-world testing with physical vehicles remains important, as our empirical study on a scaled vehicle shows that some DNN failures can be exposed only in real-world conditions. 

Our empirical results also indicate that visual odometry can be used to obtain accurate telemetry data from real SDCs, making the physical environment comparably data-rich and informative as the simulated environment. We found that the quality of driving metrics like XTE tends to be worse when collected from a real SDC, probably due to the sources of noise (e.g., illumination changes) and the stochastic effects of mechanical components (e.g., mechanical latency, friction), which are difficult to simulate accurately in the virtual environment. We also found a practical way to reduce the cost of real-world testing, based on uncertainty measures taken in simulation: successful test scenarios should be prioritized by decreasing uncertainty when scheduling them for execution on a real SDC.

%% file: acks.tex
\section*{Acknowledgments} 
This work was partially supported by the H2020 project PRECRIME, funded under the ERC Advanced Grant 2017 Program (ERC Grant Agreement n. 787703). We thank Roberto Minelli for the engineering advice and support concerning the \dk.